\documentclass{PoS}

\newcommand\figwidth{0.66}

\title{Localization and topology in high temperature QCD}

\ShortTitle{Localization and topology in high temperature QCD}

\author{\speaker{Tamas G.\ Kovacs}\thanks{Supported by OTKA Hungarian Science
    Fund under Grant No. OTKA-K-113034.}\\
        Institute for Nuclear Research, Debrecen, Hungary\\
        E-mail: \email{kgt@atomki.mta.hu}}

\author{Reka A.\ Vig$^\dagger$\\
        University of Debrecen, Hungary\\
        E-mail: \email{vig.reka@atomki.mta.hu}}

\abstract{At high temperature part of the spectrum of the quark Dirac operator
  is known to consist of localized states. This comes about because around the
  crossover temperature to the quark-gluon plasma localized states start to
  appear at the low end of the spectrum and as the system is further heated,
  states higher up in the spectrum also get localized. Since localization and
  the crossover to the chirally restored phase happen around the same
  temperature, the question of how the two phenomena are connected naturally
  arises. Here we speculate on the nature of possible gauge configurations
  that could support localized quark eigenmodes. In particular, by analyzing
  eigenmodes of the staggerd and overlap Dirac operator we show that the
  dilute gas of calorons in the high temperature phase is very unlikely to
  play a major role in localization.}

\FullConference{The 36th Annual International Symposium on Lattice Field
  Theory - LATTICE2018\\ 
		22-28 July, 2018\\
		Michigan State University, East Lansing, Michigan, USA.}

\begin{document}

\section{Introduction}

At low temperature the eigenmodes of the Dirac operator are well known to be
delocalized and in the epsilon regime the corresponding statistics of the low
eigenvalues is described by Random Matrix Theory
\cite{Verbaarschot:2000dy}. It is by now also well established that this
picture drastically changes at the crossover to the quark-gluon plasma
state. Around the crossover the lowest eigenmodes become localized and if the
temperature is further raised, the mobility edge, separating the localized
modes from the delocalized ones moves up in the spectrum, further away from
zero. How can we characterize the type of gauge field configurations that
support such localized modes?

By turning the question around we could also ask what sort of gauge field
configurations are responsible for the delocalized low Dirac modes below the
crossover temperature. A possible answer is given by the instanton liquid
model (for a review see e.g.\ \cite{Schafer:1996wv}). According to the
instanton liquid model the QCD vacuum is populated by a dense liquid of
instantons and antiinstantons, each carrying topological charge of unit
magnitude. According to the index theorem an instanton solution of the
Yang-Mills equations is always accompanied by an exact zero eigenvalue of the
corresponding covariant Dirac operator \cite{Atiyah:1968mp}. Since the
instanton-antiinstanton liquid is not an exact solution, the zero modes are
only approximate. According to the instanton liquid model the mixing of the
approximate zero modes produces the so called zero mode zone in the Dirac
spectrum, a finite density of low modes around zero virtuality, responsible
for the breaking of chiral symmetry through the Banks-Casher relation
\cite{Banks:1979yr}.

Let us consider what happens to this system when its temperature is raised. In
the Euclidean setup, as the temperature increases, the temporal extension of
the system decreases and when its linear dimension reaches the diameter of
typical instantons, they are gradually squeezed out of the box. As a result,
the instanton density and the topological susceptibility falls sharply across
the crossover, leaving only a dilute instanton gas on the quark-gluon plasma
side of the transition. In the end, the approximate zero modes of the
remaining instantons (calorons) are spatially too far away and cannot mix
sufficiently to support a finite spectral density at zero virtuality, and
chiral symmetry is restored.

Since in this way instantons play a crucial role in chiral symmetry breaking
and restoration, it is natural to expect that they are also related to the
localization transition. Indeed, the idea that instantons might play a role in
localization is almost as old as that of localization in QCD
\cite{GarciaGarcia:2005vj}. However, since zero modes are especially sensitive
to chiral properties of the Dirac operator, for a direct study of the
relationship between localization and the zero mode zone in the Dirac spectrum
on the lattice, a chirally symmetric Dirac operator is needed. In the present
paper we use staggered 2-stout \cite{Aoki:2005vt} and overlap quarks
\cite{Narayanan:1993sk} on quenched SU(3) gauge backgrounds to study the
connection between localization and the zero mode zone. It was already
observed in the early days of the overlap that on quenched gauge
configurations above $T_c$ the zero mode zone can be identified as a ``bump''
near zero virtuality in the overlap spectral density
\cite{Edwards:1999zm}. Here we show that stout smearing and finer lattices
allow a similar identification also in the staggered Dirac spectrum. Our main
result is that localized modes extend well beyond the zero mode zone and for
increasing temperatures a smaller and smaller fraction of localized modes is
contained in the zero mode zone. Therefore, a simple understanding of
localized modes as non-mixing approximate topological zero modes is not
possible.

\section{The zero mode zone}

In this section we show how the zero mode zone can be identified and
distinguished from the rest of the Dirac spectrum. This is a nontrivial task
since at low temperature both the spectral density and other properties of the
eigenmodes change monotonically as we go up in the spectrum starting from
zero. This is not unexpected since at low temperature the instanton liquid is
dense, the distance between neighboring topological objects is comparable to
their size. It is known that in the field of an instanton-antiinstanton pair
the two would be zero modes split into a pair of complex conjugate Dirac
eigenvalues and their splitting is controlled by the distance (compared to
their size) of the two topological objects\footnote{The splitting also depends
  on the relative gauge orientation of the instanton and antiinstanton, but
  this is not relevant to the present argument.}\cite{Schafer:1996wv}; the
closer the instanton and the antiinstanton are, the larger the splitting
is. Therefore, in a densely packed medium, the zero mode zone extends
higher up in the spectrum and there is no well defined boundary between the
zero mode zone and the bulk of the spectrum. 

This situation, however, changes drastically through the crossover where the
instanton density falls sharply. At high temperature the instanton gas becomes
more and more dilute and also the typical instanton size decreases, as larger
instantons are squeezed out of the system by the decreasing temporal size. As
a result, the typical distance between neighboring topological objects becomes
much larger than their size, which in turn will produce smaller splittings for
the mixing approximate zero modes.

However, to resolve these fine details of the spectrum, one needs a lattice
Dirac operator which at least approximately respects chiral symmetry. This is
because a lattice Dirac operator that explicitly breaks chiral symmetry does
not have zero modes, the magnitude of its would be zero eigenvalues is
controlled by the explicit chiral symmetry breaking of the operator. Thus, if
the Dirac operator has bad chiral properties, the splittings in the zero mode
zone are not governed by the physics of the instanton gas but by the explicit
chiral symmetry breaking of the Dirac operator, a lattice artifact that will
eventually disappear in the continuum limit.

\begin{figure}
\begin{center}
\includegraphics[width=\figwidth\columnwidth,keepaspectratio]{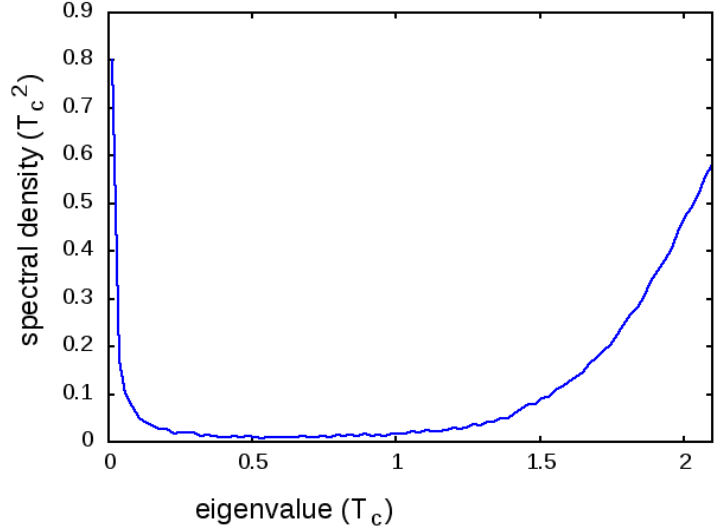}
\caption{\label{fig:spd6} The spectral density (normalized by the 3d spatial
  volume) of the overlap Dirac operator on quenched lattices with $N_t=6$ at a
  temperature of $T=1.06 T_c$. The units on the axes are appropriate powers of
  $T_c$. }
\end{center}
\end{figure}

For this reason we start our study by using the overlap Dirac operator that
has an exact chiral symmetry and consequently exact topological zero modes. In
Fig.\ \ref{fig:spd6} we plot the spectral density of the overlap Dirac
operator on a quenched ensemble of lattices with $N_t=6$, just above the
critical temperature. In contrast to the low temperature case where the
spectral density increases monotonically, here we see an accumulation of
eigenvalues near zero, followed by a dip and then a rising spectral density as
we enter the bulk of the spectrum. We emphasize that the peak near zero is not
due to topological zero modes. Since the overlap has exact zero modes, those
would show up as a trivial delta peak at zero, and we removed that from the
plot. A similar accumulation of near zero modes was previously observed in
Ref.\ \cite{Edwards:1999zm}, albeit on coarser lattices.

Since this low part of the spectrum clearly separates from the bulk, it is a
natural candidate for the zero mode zone. The question is whether it really
contains only topology related approximate zero modes and whether it contains
all of them. Fortunately, the overlap spectrum provides a simple way of
independently estimating the expected number of topological objects and their
approximate zero modes. By the index theorem, on each configuration the number
of zero modes gives the net topological charge\footnote{In principle the
  difference of the number of negative and positive chirality zero modes is
  equal to the topological charge. However, in practice one never encounters
  zero modes of opposite chirality on the same configuration.} which in turn
can be used to estimate the topological susceptibility. If we {\it assume}
that above $T_c$ the instanton gas is dilute enough so that interactions among
topological objects can be neglected, then topological objects occur
independently. In this case a simple calculation shows that the density of
topological objects (i.e.\ the instanton plus antiinstanton density) is equal
to the susceptibility. Thus the susceptibility gives an independent estimate
of the expected number of eigenmodes in the zero mode zone. 

\begin{figure}
\begin{center}
\includegraphics[width=\figwidth\columnwidth,keepaspectratio]{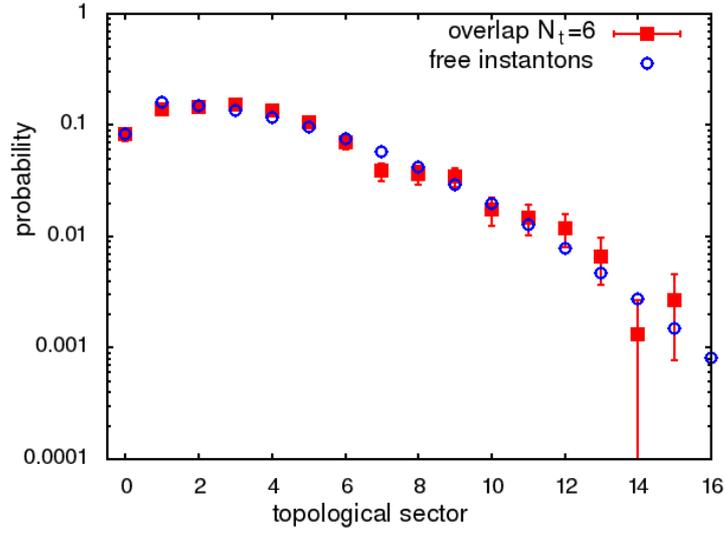}
\caption{\label{fig:qdist} The probability of different topological charge
  sectors in the lattice simulation at $T=1.06T_c$ and the corresponding
  distribution in a free instanton antiinstanton gas with the same
  topological susceptibility. Since the distribution is expected to be
  symmetric, the probabilities of positive and negative charges of the same
  magnitude have been added.}
\end{center}
\end{figure}

A key assumption of the above line of argument is that the interaction among
topological objects can be neglected, they occur independently. This can be
tested by comparing the distribution of the topological charge found in
lattice simulations to that of the free instanton gas with the same
topological susceptibility. In Fig.\ \ref{fig:qdist} we compare the two
distributions and find good agreement. 

\begin{figure}
\begin{center}
\includegraphics[width=\figwidth\columnwidth,keepaspectratio]{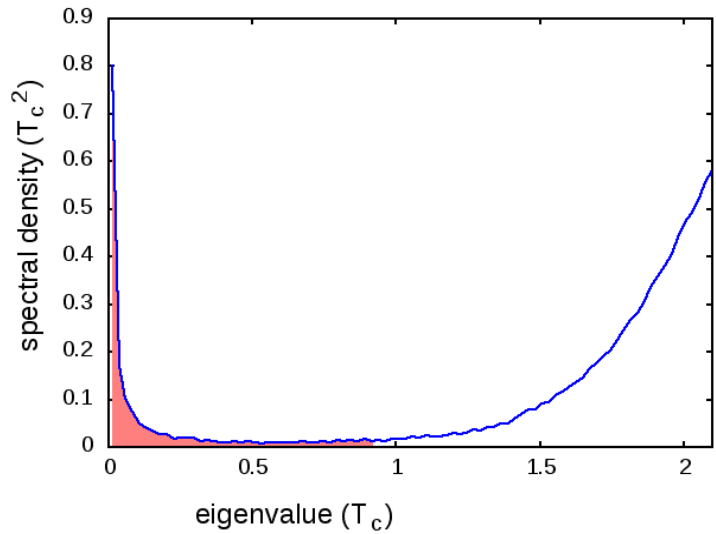}
\caption{\label{fig:spd_zmz} The spectral density of the overlap Dirac
  operator on quenched lattices with $N_t=6$ at a temperature of $T=1.06
  T_c$. The units on the axes are appropriate powers of $T_c$. The red shaded
  region shows how far up in the spectrum the zero mode zone extends. This is
  estimated based on the topological susceptibility, as given by the number of
  zero modes of the overlap Dirac operator.}
\end{center}
\end{figure}

Having justified that our assumption about the independence of topological
objects is valid, based on the topological susceptibility we can estimate how
far up in the spectrum the zero mode zone extends. This is shown in
Fig.\ \ref{fig:spd_zmz}, where we plot the overlap spectral density with the
red shaded region indicating the extension of the zero mode zone in the
spectrum. We also performed simulations at a number of other temperatures
ranging from $1.02 T_c$ to $1.12 T_c$ and the picture is qualitatively similar
everywhere. This strongly suggests that the peak in the spectral density
indeed represents the zero mode zone.

\section{The zero mode zone and the mobility edge}

\begin{figure}
\begin{center}
\includegraphics[width=\figwidth\columnwidth,keepaspectratio]{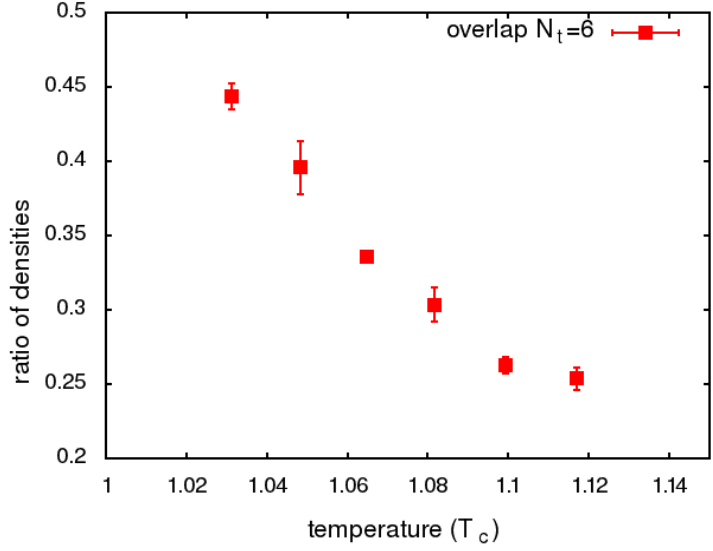}
\caption{\label{fig:top_p_loc_ov} The ratio of the number of modes in the zero
mode zone and the number of localized modes as a function of the temperature
(in units of $T_c$).}
\end{center}
\end{figure}

A simple explanation for the appearance of localized modes above $T_c$ would
be that the localized modes are nothing else but the modes in the zero mode
zone that get separated from the bulk of the spectrum, as we saw
above. Localized modes at the low end of the spectrum can be easily
distinguished from the higher, delocalized part of the spectrum by tracing how
the spectral statistics changes along the spectrum. For a detailed account of
the determination of the mobility edge, separating localized and delocalized
modes, see the contribution by R.\ A.\ Vig at this conference and also
\cite{Kovacs:2017uiz}. We can now count the average number of eigenvalues
below the mobility edge and compare that to the number of eigenvalues in the
zero mode zone to see whether the zero mode zone can account for
localization. In Fig.\ \ref{fig:top_p_loc_ov} we plot the ratio of the number
of modes in the zero mode zone to the number of localized modes as a function
of the temperature. This shows that even right above $T_c$ not more than 50\%
of the localized modes are contained in the zero mode zone and this fraction
falls to 25\% already at $T=1.12T_c$.

\begin{figure}
\begin{center}
\includegraphics[width=\figwidth\columnwidth,keepaspectratio]{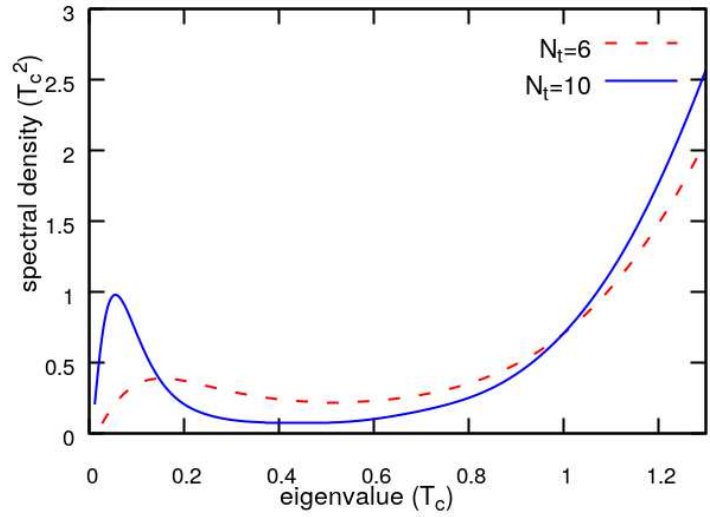}
\caption{\label{fig:spd_ks} The spectral density of the 2-stout staggered
  Dirac operator on $N_t=6$ and $N_t=10$ lattices at the same physical
  temperature $T=1.06 T_c$.}
\end{center}
\end{figure}

Our limited data set for the overlap spectra does not allow us to follow how
this ratio continues to change as the temperature increases further. However,
this can be qualitatively seen using our more extensive set of staggered Dirac
spectra. Indeed, on finer lattices the spectral density of our 2-stout smeared
staggered Dirac operator shows features similar to those of the overlap and
this allows the approximate identification of the zero mode zone here as
well. We demonstrate this in Fig.\ \ref{fig:spd_ks} by showing the spectral
density of the staggered operator on $N_t=6$ and $N_t=10$ lattices at the same
physical temperature, $T=1.06T_c$. It is easily seen that as the lattice
becomes finer, the bump near zero becomes more pronounced and more separated
from the bulk of the spectrum. We note that as the staggered operator does not
provide unambiguous information about the topological susceptibility, the
staggered zero mode zone can only be approximately identified using the
location of the bump. However, this identification becomes increasingly
precise in the continuum limit.

\begin{figure}
\begin{center}
\includegraphics[width=\figwidth\columnwidth,keepaspectratio]{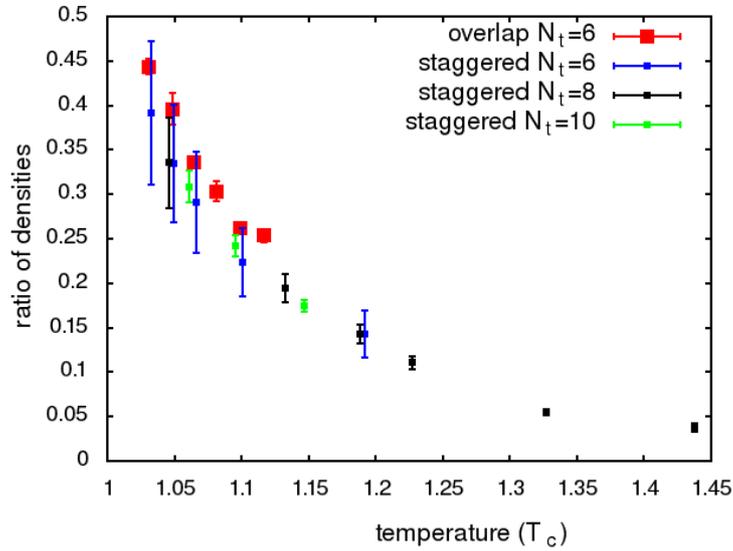}
\caption{\label{fig:top_p_loc_all} The ratio of the number of modes in the zero
mode zone and the number of localized modes as a function of the temperature
(in units of $T_c$). Here we show data based on the 2-stout staggered Dirac
operator along with the overlap data in the previously presented overlap data.}
\end{center}
\end{figure}

Keeping this in mind, in Fig.\ \ref{fig:top_p_loc_all} we present staggered
data for the ratio of the size of the zero mode zone and the localized region
in the spectrum. The staggered data shown corresponds to three different
lattice spacings ($N_t=6,8,10$) and for comparison we included also the
previously presented overlap data. The error bars for the staggered data also
contain the uncertainty in estimating the size of the zero mode zone. Note
that due to the less expressed nature of the bump in the spectral density,
this uncertainty is larger for the coarser lattices. This is also the reason
why the error bars are decreasing towards higher temperatures since for a
given $N_t$ the temperature is set by changing the lattice spacing (the gauge
coupling) and higher temperatures correspond to finer lattices.

The staggered data obtained at different lattice spacings are consistent and
they also agree with the overlap data. This shows that our results can be
safely considered to be a good qualitative estimate of the continuum limit. As
the temperature increases, the ratio keeps falling and around $1.5 T_c$ only
about 5\% of the localized modes is contained in the zero mode zone. This
clearly rules out a simple identification of the zero mode zone with the
region of localized modes, and makes a direct connection between localization
and topology rather unlikely. 

\section{Conclusions}

In the present paper we studied the question of whether the zero mode zone of
the high temperature QCD Dirac operator can be identified with the region of
localized modes at the low end of the spectrum. We showed that both with the
overlap operator and the 2-stout staggered operator on fine enough lattices
the zero mode zone can be identified as a bump in the spectral density near
zero. However, the zero mode zone turns out to contain only a fraction of the
localized modes and this fraction falls sharply at higher
temperatures. Therefore, a simple identification of localized modes with
approximate zero modes of topological origin is ruled out. Finally, we note
that although the present study is based on the quenched approximation, we do
not expect that the inclusion of dynamical quarks would change our main
conclusion. Indeed, dynamical quarks suppress fluctuations of the topological
charge and in the presence of dynamical quarks, topology is even less likely to
provide an explanation for localization.


\begin{thebibliography}{99}

\bibitem{Verbaarschot:2000dy} 
  J.~J.~M.~Verbaarschot and T.~Wettig,
  Ann.\ Rev.\ Nucl.\ Part.\ Sci.\  {\bf 50}, 343 (2000)
  [hep-ph/0003017].

\bibitem{Schafer:1996wv} 
  T.~Sch\"afer and E.~V.~Shuryak,
  Rev.\ Mod.\ Phys.\  {\bf 70}, 323 (1998)
  [hep-ph/9610451].

\bibitem{Atiyah:1968mp} 
  M.~F.~Atiyah and I.~M.~Singer,
  Annals Math.\  {\bf 87}, 484 (1968).

\bibitem{Banks:1979yr} 
  T.~Banks and A.~Casher,
  Nucl.\ Phys.\ B {\bf 169}, 103 (1980).

\bibitem{GarciaGarcia:2005vj} 
  A.~M.~Garcia-Garcia and J.~C.~Osborn,
  Nucl.\ Phys.\ A {\bf 770}, 141 (2006)
  [hep-lat/0512025];
     %
  A.~M.~Garcia-Garcia and J.~C.~Osborn,
  Phys.\ Rev.\ D {\bf 75}, 034503 (2007)
  [hep-lat/0611019].

\bibitem{Aoki:2005vt} 
  Y.~Aoki, Z.~Fodor, S.~D.~Katz and K.~K.~Szabo,
  JHEP {\bf 0601}, 089 (2006)
  [hep-lat/0510084].

\bibitem{Narayanan:1993sk} 
  R.~Narayanan and H.~Neuberger,
  Nucl.\ Phys.\ B {\bf 412}, 574 (1994)
  [hep-lat/9307006];
     %
  R.~Narayanan and H.~Neuberger,
  Nucl.\ Phys.\ B {\bf 443}, 305 (1995)
  [hep-th/9411108].

\bibitem{Edwards:1999zm} 
  R.~G.~Edwards, U.~M.~Heller, J.~E.~Kiskis and R.~Narayanan,
  Phys.\ Rev.\ D {\bf 61}, 074504 (2000)
  [hep-lat/9910041].

\bibitem{Kovacs:2017uiz} 
  T.~G.~Kovacs and R.~A.~Vig,
  Phys.\ Rev.\ D {\bf 97}, no. 1, 014502 (2018)
  [arXiv:1706.03562 [hep-lat]].

\end{thebibliography}
\end{document}